\documentclass{Interspeech}
\usepackage{subcaption}
\usepackage{url}
\PassOptionsToPackage{hyphens}{url}\usepackage{hyperref}
\usepackage{multirow}
\usepackage{siunitx}
\usepackage{booktabs}

\usepackage[absolute,showboxes]{textpos}
\usepackage[symbol]{footmisc}
\renewcommand*{\thefootnote}{\fnsymbol{footnote}}

\interspeechcameraready

\title{Unified Microphone Conversion: Many-to-Many Device Mapping via Feature-wise Linear Modulation}

\author[affiliation={1,2}]{$^\star$Myeonghoon}{Ryu}
\author[affiliation={3}]{$^\star$Hongseok}{Oh}
\author[affiliation={1}]{Suji}{Lee}
\author[affiliation={1}]{Han}{Park}

\affiliation{}{Deeply Inc.}{Republic of Korea}
\affiliation{}{Seoul National University}{Republic of Korea}
\affiliation{}{University of California, San Diego}{USA}
\email{myeonghoon.ryu@deeply.co.kr, h1oh@ucsd.edu, suji@deeply.co.kr, han@deeply.co.kr}

\usepackage{comment}

\begin{document}
\maketitle
\begingroup
\renewcommand\thefootnote{}\footnotetext{$^\star$ These authors contributed equally to this work.}
\endgroup
\begin{abstract}
    We present Unified Microphone Conversion, a unified generative framework designed to bolster sound event classification (SEC) systems against device variability. While our prior CycleGAN-based methods effectively simulate device characteristics, they require separate models for each device pair, limiting scalability. Our approach overcomes this constraint by conditioning the generator on frequency response data, enabling many-to-many device mappings through unpaired training. We integrate frequency-response information via Feature-wise Linear Modulation, further enhancing scalability. Additionally, incorporating synthetic frequency response differences improves the applicability of our framework for real-world application. Experimental results show that our method outperforms the state-of-the-art by 2.6\% and reduces variability by 0.8\% in macro-average F1 score.
\end{abstract}
\section{Introduction}
\label{sec:intro}
Sound Event Classification (SEC) involves identifying audio events in a sound recording, enabling systems to recognize specific sounds such as speech, music, or environmental noises. However, the accuracy of these systems is often compromised by distortions introduced by recording devices. These distortions can arise from device-specific characteristics—such as frequency response mismatches, microphone sensitivity, or built-in signal processing algorithms—as well as environmental factors, including background noise and reverberation. Although often unnoticed by human listeners, these distortions can significantly diminish the accuracy of SEC systems. \cite{Martínmorato2022lowcomplexity}

Previous methods to address the impact of device variability have mainly focused on data augmentation and data-independent normalization techniques. \cite{Schmid2022, kim22_interspeech, 9747680} Our previous method tackles this challenge by utilizing a CycleGAN \cite{zhu2017unpaired} to generate synthetic training audio samples, thereby simulating recordings captured with various devices. While this approach successfully leverages unpaired training data to model device characteristics, it still relies on a deterministic, one-to-one mapping between two specific device domains. Consequently, a separate model must be trained for each device pair, leading to scalability challenges when dealing with multiple devices.

Motivated by this limitation, we propose a many-to-many device mapping approach utilizing a CycleGAN framework combined with Feature-wise Linear Modulation (FiLM) \cite{perez2018film}, incorporating frequency response data of recording devices. We hypothesize that integrating device frequency response information via FiLM can accurately specify the desired inter-domain mappings while maintaining consistent acoustic information. Specifically, we modulate the channel-wise statistics of the generator’s intermediate embeddings by incorporating frequency-response-related embeddings, allowing the model to capture inter-device transformations. By not restricting the model to particular source–target pairs, this method provides a scalable and adaptable solution to address device variability in sound event classification. Moreover, we introduce a synthetic frequency difference generation algorithm to enhance the scalability of the Unified Microphone Conversion network, enabling more efficient training of SEC models across diverse domains without needing additional frequency response data. 

\section{Related Work}
\label{related_work}
In tackling device variability, various strategies have been developed through the DCASE Challenges \cite{Martínmorato2022lowcomplexity}. Common data augmentation techniques such as noise addition, convolving room impulse response, pitch shifting, SpecAugment \cite{Park2019}, and MixUp \cite{zhang2018mixup} are frequently employed to diversify training data with a range of acoustic conditions. However, while these methods can provide marginal performance gains, they often fail to account for the distortions that arise from recording devices. Consequently, the mismatch in frequency response or built-in signal processing among devices remains unaddressed, limiting the effectiveness of these methods in real-world scenarios.

More complex methods that manipulate frequency statistics have emerged to enhance generalization in the presence of device variability. For example, MixStyle \cite{Schmid2022} exchanges frequency components to mimic the effects of different devices. Likewise, Residual Normalization \cite{Kim2021b} and Relaxed Instance Frequency-wise Normalization (RFN) \cite{kim22_interspeech} focus on normalizing frequency-wise statistics to counteract device-specific distortions. Furthermore, FilterAugment \cite{Nam2021} simulates device-induced filtering by applying weights to frequency bands, helping the model handle device-specific spectral variations. 

\begin{figure*}[th]
  \centering
  \includegraphics[width=0.8\linewidth]{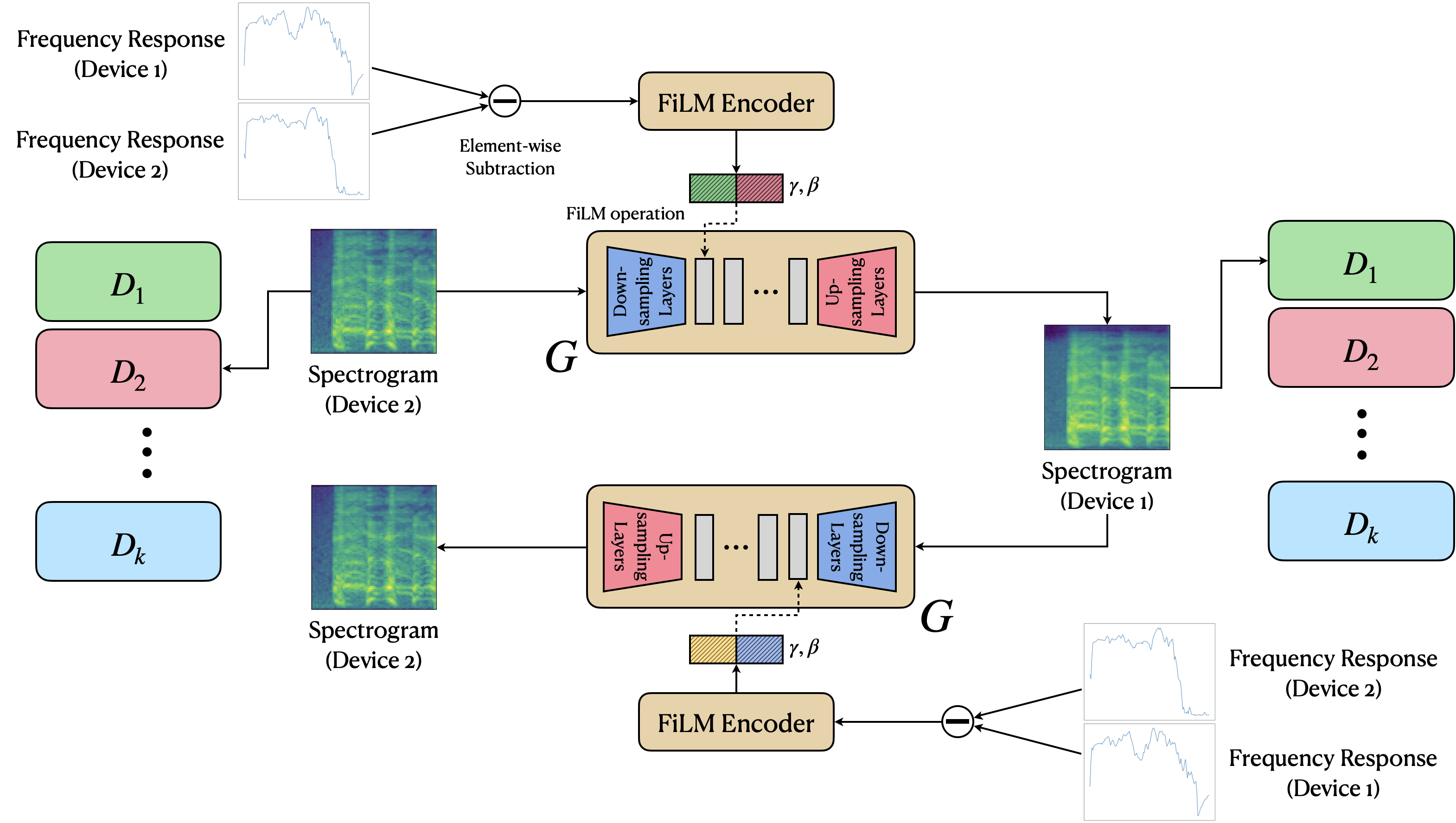}
  \caption{
  The training pipeline of the Unified Microphone Conversion model. The model receives an input spectrogram and a frequency response difference—computed by subtracting the frequency response of the source device from that of the target device—and outputs a spectrogram that simulates the target device domain. The FiLM encoder is jointly trained with $G$ to learn device-specific modulation parameters, while each domain $i$ is assigned its own discriminator $D_i$ to enforce alignment with the target device’s distribution.}
  \label{fig:umc_training}
\end{figure*}

Recently, a Microphone Conversion approach \cite{10447432} utilized the CycleGAN framework to address device variability by generating synthetic spectrograms that mimic recordings from various devices. While it outperformed all previous methods and established the state-of-the-art for handling device variability across multiple devices, its reliance on one-to-one domain mappings limits scalability, as a separate model must be trained for each device pair—making it a strong but impractical baseline for comparison. This requirement becomes increasingly challenging with the expanding array of recording devices, especially if new data becomes available or additional devices are introduced, as it necessitates retraining the network. 

\begin{figure}[t]
  \centering
  \includegraphics[width=\linewidth]{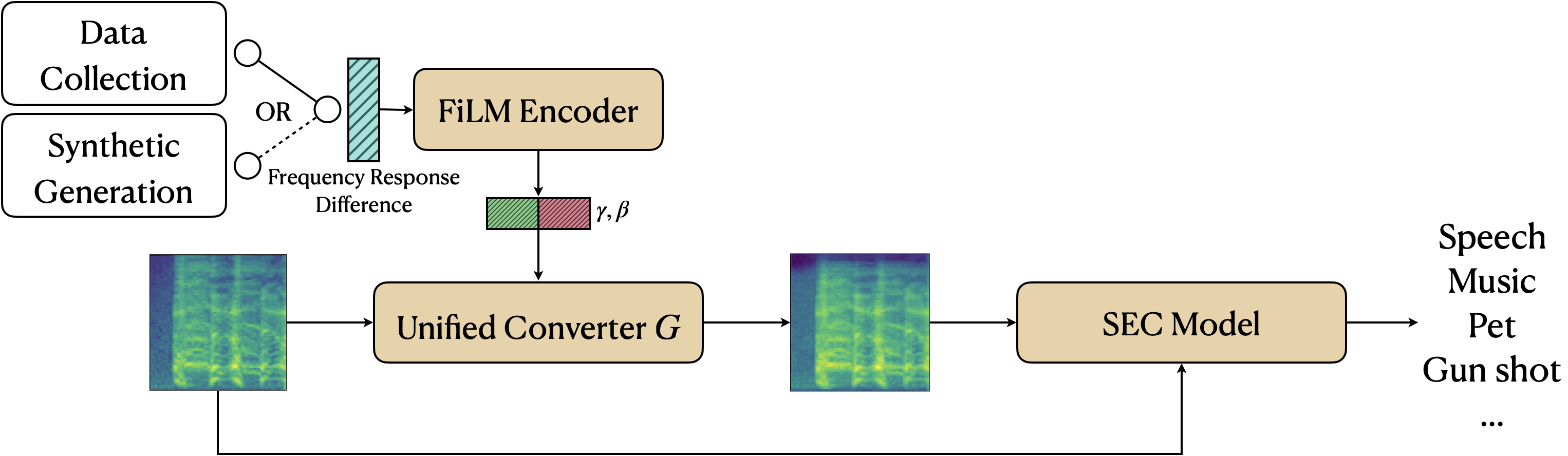}
  \caption{
  The training phase for SEC models using the trained Unified Microphone Conversion model $G$. Either real-world or synthetic frequency response difference data are provided to the FiLM encoder, which then modulates the generator $G$, producing converted spectrograms for robust SEC model training.}
  \label{fig:diagram}
\end{figure}

\begin{figure}[t]
  \centering
  \includegraphics[width=\linewidth]{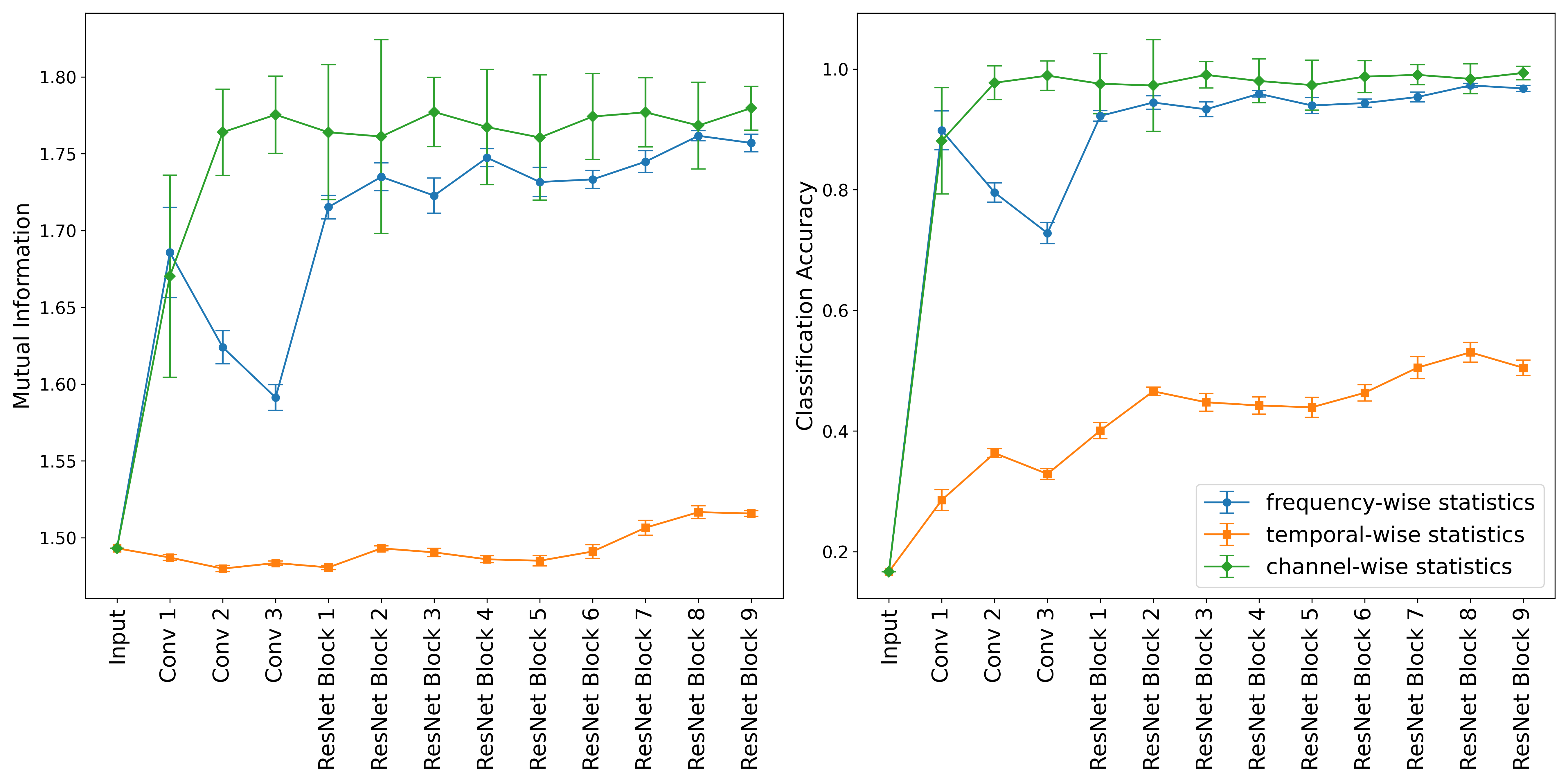}
  \caption{(left) Mutual information estimates between the target device and the dimension-wise statistics—the embeddings averaged across each dimension—at different layers of the Microphone Conversion network. (right) Classification accuracy of the target device given these dimension-wise statistics. }
  \label{fig:mutual_info}
\end{figure}

\begin{figure*}[t]
  \centering
  \includegraphics[width=\linewidth]{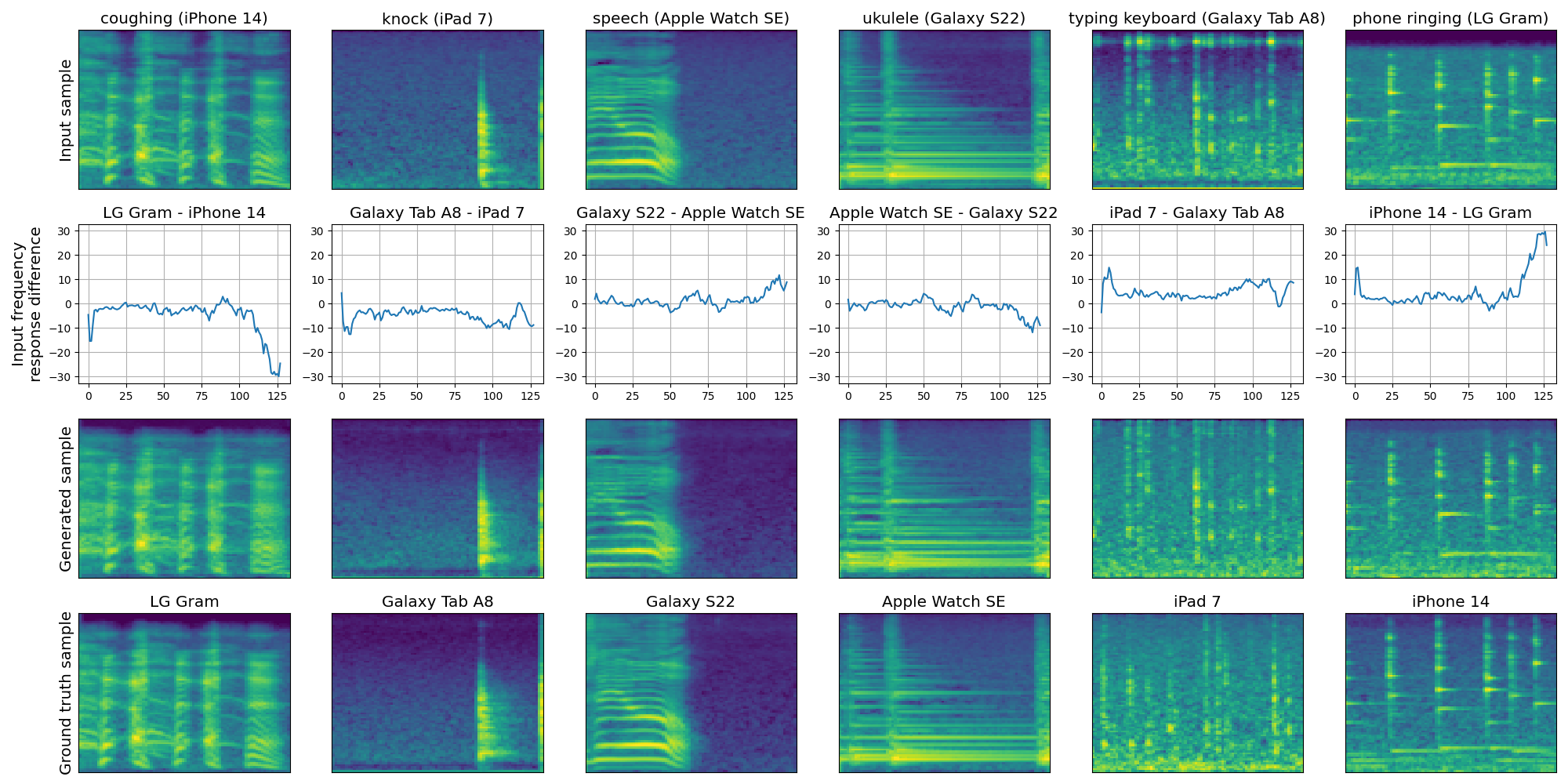}
  \caption{
  The first two rows display the input spectrogram of different acoustic contents and recording devices, and the frequency response difference between the target and input devices. The third row presents samples generated by Unified Microphone Conversion using these inputs, while the final row depicts ground truth spectrograms from the target devices.}
  \label{fig:source_target}
\end{figure*}

\section{Methodology}

\subsection{Unified Microphone Conversion}
We aim to generate a spectrogram corresponding to a different recording device, given the original spectrogram and the frequency response difference between the input and target devices. This approach eliminates the need for multiple generators for each device pair, allowing more scalable architecture.

In this work, we propose a Unified Microphone Conversion framework that combines key concepts from CycleGAN and FiLM to tackle device variability in audio recordings. Our method conditions the CycleGAN generator with FiLM encoder utilizing the frequency response difference, granting more versatile and efficient handling of diverse devices, overcoming the limitation of CycleGAN, which relies on a bijective mapping.

In contrast to the conventional CycleGAN setup, which employs two generators and two discriminators, our modified architecture consolidates two generators into a single unified generator augmented by outputs of a FiLM encoder. This generator is supported by $k$ discriminators, each assigned to one of $k$ distinct device domains. We adopt the architectures for the generator and discriminator from Ryu et al. \cite{10447432}. 

\begin{table*}[t]
\caption{Macro-average F1 scores of SEC models trained on sound samples from each source device with the Unified Microphone Conversion network. The performance is evaluated utilizing validation sound samples from all seven devices.}
\label{tab:result}
\begin{minipage}[t]{\linewidth}
\subcaption{Experiment result of Unified-MC-Real with recorded frequency response}
\label{tab:result_real}
\begin{tabular}{lccccccc}
\toprule
\multicolumn{1}{l}{\multirow{2}{*}{Source Device}} & \multicolumn{7}{c}{Target Device}                                                        \\\cmidrule{2-8}
\multicolumn{1}{l}{}                  & iPhone 14 & Galaxy S22 & iPad 7 & Galaxy Tab A8 & Apple Watch SE & MacBook Pro & LG Gram \\ \midrule
iPhone14 &0.974 & 0.968 & 0.936 & 0.938 & 0.883 & 0.912 & 0.917\\
GalaxyS22 &0.964 & 0.966 & 0.920 & 0.903 & 0.888 & 0.916 & 0.879\\
iPad7 &0.945 & 0.928 & 0.964 & 0.928 & 0.903 & 0.863 & 0.904\\
GalaxyTabA8 &0.953 & 0.940 & 0.930 & 0.973 & 0.868 & 0.802 & 0.869\\
AppleWatchSE &0.894 & 0.882 & 0.874 & 0.795 & 0.967 & 0.907 & 0.735\\
MacbookPro &0.918 & 0.922 & 0.862 & 0.741 & 0.914 & 0.977 & 0.852\\
LG-Gram &0.907 & 0.913 & 0.878 & 0.809 & 0.823 & 0.849 & 0.974\\\bottomrule
\end{tabular}
\end{minipage}

\begin{minipage}[t]{\linewidth}
\subcaption{Experiment result of Unified-MC-Synth with synthetic frequency response}
\label{tab:result_synth}
\begin{tabular}{lccccccc}
\toprule
\multicolumn{1}{l}{\multirow{2}{*}{Source Device}} & \multicolumn{7}{c}{Target Device}                                                        \\\cmidrule{2-8}
\multicolumn{1}{l}{}                  & iPhone 14 & Galaxy S22 & iPad 7 & Galaxy Tab A8 & Apple Watch SE & MacBook Pro & LG Gram \\ \midrule
iPhone14 &0.974 & 0.962 & 0.943 & 0.928 & 0.910 & 0.914 & 0.901\\
GalaxyS22 &0.945 & 0.970 & 0.881 & 0.878 & 0.878 & 0.905 & 0.858\\
iPad7 &0.950 & 0.929 & 0.964 & 0.909 & 0.902 & 0.869 & 0.890\\
GalaxyTabA8 &0.942 & 0.922 & 0.906 & 0.972 & 0.837 & 0.801 & 0.818\\
AppleWatchSE &0.896 & 0.865 & 0.860 & 0.752 & 0.970 & 0.891 & 0.764\\
MacbookPro &0.928 & 0.908 & 0.870 & 0.786 & 0.913 & 0.973 & 0.858\\
LG-Gram &0.813 & 0.842 & 0.791 & 0.701 & 0.668 & 0.806 & 0.975\\\bottomrule
\end{tabular}
\end{minipage}

\end{table*}

\begin{table*}[t]
\caption{Results for generalization capability of our method and previous methods on the validation set. Source device (S) is iPhone 14, and target devices (T1 - T6) are Galaxy S22, iPad 7, Galaxy Tab A8, Apple Watch SE, Macbook Pro ('20), and LG Gram ('20), respectively. The last column shows an average and 95\% confidence interval of the performance.}
\label{tab:result2}
\centering
\begin{tabular}{lcccccccc}
\toprule
\multicolumn{1}{c}{\multirow{2}{*}{Method}} & \multicolumn{8}{c}{F1 Score}                                                                                     \\ \cmidrule{2-9} 
\multicolumn{1}{c}{}                        & S & T1 & T2 & T3 & T4 & T5 & T6 & Overall (- S) \\ \midrule
Baseline                                    & 0.982     & 0.409      & 0.709  & 0.248         & 0.471          & 0.687       & 0.491   & 0.503 ± 0.167         \\
MC-100-Gen                          & 0.981     & 0.958      & 0.912  & 0.894         & 0.899          & 0.831       & 0.852   & 0.891 ± 0.043         \\
MC-200-Gen                          & 0.982     & \textbf{0.969}      & 0.909  & 0.903         & \textbf{0.912}          & 0.859       & 0.887   & 0.907 ± 0.035         \\
Unified-MC-Real      &  0.975 & 0.967 & \textbf{0.936} & \textbf{0.939} & 0.885 & \textbf{0.913} & 0.913 & \textbf{0.933 ± 0.027}\\
Unified-MC-Synth      &  0.974 & 0.968 & \textbf{0.936} & 0.938 & 0.883 & 0.912 & \textbf{0.917} & \textbf{0.933 ± 0.027}\\\midrule
Ideal                                        & 0.983     & 0.982      & 0.972  & 0.985         & 0.979          & 0.983       & 0.986   & 0.981 ± 0.005         \\ \bottomrule
\end{tabular}
\end{table*}
%



\subsection{FiLM Encoder}
A FiLM encoder maps the frequency response difference to the scaling and shifting factors ($\gamma$, $\beta$) for each feature map. We apply FiLM-based modulation to the channel-wise feature statistics within the first residual block of the Unified Microphone Conversion network, harnessing scaling and shifting factors generated by our FiLM encoder. As illustrated in Figure \ref{fig:mutual_info}, this choice is motivated by an analysis showing that channel-wise statistics exhibit the highest mutual information with the target device. The mutual information is estimated following the method presented by Wang et al. \cite{wang2021revisiting} The FiLM encoder itself comprises three convolutional blocks—each consisting of a 1D convolution, instance normalization, a ReLU activation, and a multi-layer perceptron that outputs the modulation factors.

\subsection{Synthetic Frequency Response Difference}
We propose a method to randomly generate synthetic frequency response differences for the FiLM encoder during the inference phase of the Unified Microphone Conversion, which is the training phase of the SEC model. This approach mitigates the high cost and technical complexity of collecting frequency response data, while producing more diverse output spectrograms.

To achieve this, we first segment the frequency range into multiple bands and assign random reference values at the boundaries of these bands. These values are samples within a predefined range from a uniform distribution and randomly assigned positive or negative signs to introduce variation. To ensure a gradual transition across the frequency bins, linear interpolation is applied between adjacent boundary values, forming a continuous frequency response difference curve. This method effectively simulates diverse frequency response variations without requiring real-world calibration data, making it practical for augmenting training datasets for SEC models.

%

\section{Experiments and Results}

\subsection{Dataset}
\label{section:dataset}
The development set comprises 75 unique sound events and 315,966 audio segments (each 930 ms), recorded using seven end-user devices (iPhone 14, Galaxy S22, iPad 7, Galaxy Tab A8, Apple Watch SE, MacBook Pro 2020, and LG Gram 2020) in an ISO 3745–compliant anechoic chamber. \cite{iso3745} The audio is downsampled to 22,050 Hz and transformed into log Mel spectrograms using a 1,024-bin Hanning window, a 256-bin hop length, and 80 Mel bands. Following the procedure described in Sections 2 and 3.2 of Ryu et al. \cite{10447432}, we adopt the same dataset split for a fair evaluation—$data_{train,mc}$, $data_{train,sec}$, and $data_{val,sec}$—for Unified Microphone Conversion training, SEC model training, and SEC model validation, respectively.

We extend the dataset by adding frequency response data of each device, defined as log Mel spectra (128 Mel bands) from 200-ms segments of recorded impulses. The Kronecker delta function is used to generate the impulse with readily available mobile phones, providing a cost-effective alternative to more rigorous methods for characterizing the frequency response. Moreover, as the DCASE dataset lacks impulse response information, our dataset-which includes real device IRs and greater class and device diversity-is more suitable for our experiments.

\subsection{Implementation Details}
\label{section:implementation}
For the training of the Unified Microphone Conversion network, we use Adam optimizer \cite{DBLP:journals/corr/KingmaB14} for 100 epochs with a learning rate $5 \times 10^{-4}$, divided by 10 every 30 epochs, with a batch size 400 on 4 RTX 4090s. We adopt a generated image buffer \cite{shrivastava2017learning} for each discriminator. We replace the negative log likelihood loss with the least square loss \cite{mao2017least}. The cycle consistency loss weight set to 10 to emphasize its importance in the total loss function. For the SEC systems, ResNet-50 \cite{he2016residual} serve as the backbone architecture. We use the AdamW optimizer \cite{loshchilov2018decoupled} for 200 epochs with a learning rate $1 \times 10^{-3}$, divided by 10 every 25 epochs, with a batch size 100 on a RTX 4090. Each SEC system is trained exclusively on recordings from a single source device, which are converted to six other devices via the Unified Microphone Conversion, while some samples remain unaltered. Figure \ref{fig:umc_training} and \ref{fig:diagram} illustrate the training and deployment pipeline of Unified Microphone Conversion model.

We adopt two variants of the proposed approach, Unified-MC-Real and Unified-MC-Synth. Unified-MC-Real leverages measured frequency responses collected from actual devices to ensure accurate modeling of real-world characteristics. In contrast, Unified-MC-Synth employs synthetically generated frequency responses, removing the need for device-specific measurements while still capturing essential spectral variations.

\subsection{Results}
Table \ref{tab:result} presents the performance of Unified-MC-Real and Unified-MC-Synth, each trained on data from different source devices. Each row represents a model trained on a particular device, showcasing robust performance across heterogeneous recording conditions. The results confirm that both methods significantly enhance resilience against device mismatch, as evidenced by consistent performance gains in classification task.

In Table \ref{tab:result2}, the Ideal system represents an ideal scenario in which data from all seven devices are available during training. The Baseline system is a SEC model trained on iPhone 14 data only, without any additional techniques, which demonstrates the extent of performance deterioration caused by device mismatch. The state-of-the-art methods, MC-100-Gen and MC-200-Gen, employ multiple Microphone Conversion models for each device pair, each trained for 100 or 200 epochs, respectively, to generate augmented audio for SEC model training. Our methods substantially bridge the gap between the Baseline and Ideal systems, outperforming the state-of-the-art by 2.6\% and reducing variability by 0.8\% in the macro-average F1 score. Notably, Unified-MC-Synth performs comparably to Unified-MC-Real, despite not requiring recorded device frequency response.

\section{Conclusions and Limitations}
We propose Unified Microphone Conversion to address device variability in SEC systems, overcoming the bijective mapping limitation of CycleGAN while boosting scalability. By modulating the generator’s intermediate embeddings with device frequency response information, our approach further improves SEC performance across diverse devices, outperforming the state-of-the-art method that requires training multiple generators. Additionally, synthetic frequency response differences eliminate the need for collecting measured device data, further enhancing scalability. However, the synthetic frequency generation is currently based on hand-crafted rules. Another limitation is the absence of a direct evaluation metric for Unified Microphone Conversion. The current work relies on classification performance to assess the quality of domain adaptation.

\vfill\pagebreak

\bibliographystyle{IEEEtran}
\bibliography{main}

\end{document}